\documentclass[pre,twocolumn]{revtex4}
\usepackage{graphicx,amssymb}

\begin{document}
\title{Anomalous sensitivity to initial conditions and
  entropy production in standard maps: Nonextensive approach}

\author{Gar\'{\i}n F. J. A\~na\~nos$^{1,2}$, 
Fulvio Baldovin$^{1}$, 
Constantino Tsallis$^{1,3}$, 
}
\address{
$^1$Centro Brasileiro de Pesquisas F\'\i sicas,\
Rua Xavier Sigaud 150, 
22290-180 Rio de Janeiro-RJ, Brazil.\\
$^2$Departamento Acad\'emico de F\'{\i}sica, Universidad Nacional de Trujillo
Av. Juan Pablo II, s/n, Trujillo, Peru.\\
$^3$Santa Fe Institute,
1399 Hyde Park Road,
Santa Fe, New Mexico 87501,  USA.
}
\date{\today}
\begin{abstract}
We perform a throughout numerical study of the average
sensitivity to initial conditions and entropy production for
two symplectically coupled standard maps focusing on the
control-parameter region close to regularity. 
Although the system is ultimately strongly chaotic (positive
Lyapunov exponents), it first stays lengthily in
weak-chaotic regions (zero Lyapunov exponents).
We argue that the nonextensive
generalization of the classical formalism is an adequate
tool in order to get nontrivial information about this complex
phenomenon.
Within this context we analyze the relation
between the power-law sensitivity to initial conditions and the
entropy production. 

\noindent
PACS number: 05.45.-a, 05.20.-y, 05.45.Ac.
\end{abstract}
\maketitle

\section{Introduction}
\label{section_introduction}

Hamiltonian chaotic dynamics is
related to the description of irregular trajectories in
phase space. 
This
characteristic is closely linked to the instability of the
system  
and the entropy growth  (see, e.g., \cite{ott}).
Chaotic dynamics is in fact
associated with positive Lyapunov coefficients, which
corresponds, through the Pesin equality \cite{pesin},  to a 
positive Kolmogorov-Sinai entropy rate \cite{kolsinai}.  
However, many physical, biological, economical and other
complex systems exhibit more intricate situations,
associated to a phase space that reveals complicated
patterns and anomalous, not strongly chaotic,
dynamics \cite{gell_mann_01}. 
In many of these cases, the system displays an algebraic
sensitivity to initial conditions and
the use of the Boltzmann-Gibbs (BG) entropic
functional $S_{BG}\equiv-\sum_{i=1}^Wp_i\ln p_i$ 
for the definition of quantities such as the 
Kolmogorov-Sinai entropy rate only provides trivial
information.  
It has been first argued \cite{tsallis_01},  
then numerically exhibited 
\cite{Costa_Lyra_Plastino_CT,Lyra_CT,Latora_Baranger_Rapisarda_CT,Tirnakli_Ananos_CT,ernesto},   
and finally analytically shown 
\cite{baldovin_01,baldovin_02}
that under these
conditions the nonextensive generalization
\cite{tsallis_02} of the BG entropic form adequately
replaces these concepts.  
This generalization has been successfully checked for many cases
where long-range interactions, long-range microscopic memory, some kind of fractalization of the phase
space are inherent.
Indeed, in the case of one-dimensional dissipative maps,
following pioneering work \cite{grassberger_01} on asymptotic algebraic sensitivity
to initial conditions, it
has been numerically
\cite{tsallis_01,Costa_Lyra_Plastino_CT,Lyra_CT,Latora_Baranger_Rapisarda_CT,Tirnakli_Ananos_CT,ernesto}  
and analytically proved \cite{baldovin_01} that the
nonextensive formalism provides a meaningful description of
the critical points where the Lyapunov coefficient
vanishes. Moreover, through the nonextensive
entropy it is possible to prove a remarkable generalization
of the Pesin equality \cite{baldovin_02}.  

On the basis of these results, we present here a
nonextensive approach to the description of complex behaviors
associated to Hamiltonian systems that satisfy the
Kolmogorov-Arnold-Moser (KAM) requirements (see, e.g.,
\cite{zaslavsky_01}). 
The phase space consists in this case of complicated mixtures
of invariant KAM-tori and chaotic regions. 
A chaotic region is in contact with critical KAM-tori whose
Lyapunov coefficients are zero, and a chaotic orbit sticks
to those tori repeatedly with a power-law distribution of
sticking times (see, e.g., \cite{zaslavsky_02} and
references therein).
It is known that, under these conditions, one main generic
characteristic feature of Hamiltonian chaos is its
{\it nonergodicity}, due to the existence of a finite measure of
the regularity islands area. 
The set of islands is fractal, and thin
strips near the islands' boundary, which are termed boundary
layers, play a crucial role in the system dynamics. 
In addition to these processes, there are effects that arise
solely because of the high enough dimensionality of the system. 
One of these effects is Arnold diffusion \cite{arnold}, that
occurs if the number of degrees of freedom of the
Hamiltonian system is larger than two \cite{ott,zaslavsky_01}.
In general, for the usual case of many-body
short-range-interacting Hamiltonian systems, the combination
of these nonlinear dynamical phenomena typically makes the system to leave the generic situation of nonergodicity and eventually yields
{\it ergodicity}. 

Along these lines, anomalous effects for the sensitivity to
initial conditions and for the entropy production have already
been investigated within the nonextensive formalism
\cite{baldovin_03} for a paradigmatic
model that exhibits the KAM structure, namely the well known
standard map \cite{chirikov_01}. 
In the present paper we focus on two symplectically coupled
standard maps. The choice of {\it two} coupled maps is
because this is the lowest possible dimension ($d=4$) of Hamiltonian
maps where Arnold diffusion does take place, an effect that has
a fundamental impact for the macroscopic effects associated
with the dynamics of the system, since it guarantees an unique (connected) chaotic sea
(see, e.g., \cite{baldovin_04,baldovin_05}).

In Ref. \cite{latora_02} it was numerically shown
that, for low-dimensional Hamiltonian (hence symplectic) maps,
the Kolmogorov-Sinai entropy rate ${\cal K}$ coincides with the
entropy $K$ produced per unit time by the dynamical
evolution of a statistical ensemble of copies of the system
that is initially set far-from-equilibrium, i.e., $K={\cal K}$. 
More precisely, considering an ensemble of $N$ copies of the
map and a coarse graining partition of the phase space
composed by $W$ nonoverlapping (hyper)cells (with nonvanishing $d$ hypervolumes), 
at each iteration step $t$ a probability distribution 
$\{p_i\}_{i=1,2,...W}$ is defined by means of the occupation
number $N_i$ of each cell, $p_i\equiv N_i/N$
($\sum_ip_i=1$), and we have that 
\begin{equation}
K\equiv\lim_{t\to\infty}\lim_{W\to\infty}\lim_{N\to\infty}
\frac{\langle S_{BG}\rangle(t)}{t},
\end{equation}
where the average $\langle\rangle$
is taken considering the dynamical evolution of different
ensembles, all starting far-from-equilibrium, but with 
different initial conditions.
In this paper, as far-from-equilibrium initial conditions,
we will consider an ensemble of $N$ copies of the map all
randomly distributed inside a {\it single} (hyper)cell of the
partition. In this way the initial entropy is
zero, since all but one probabilities vanish.
Averages are then obtained by sampling the whole
phase space changing the position of the initial
(hyper)cell.    

On the other hand, one can consider the sensitivity to
initial conditions 
\begin{equation}
\xi({\mathbf x}(0),\delta{\mathbf x}(0), t)\equiv
\lim_{||\delta{\mathbf x}(0)||\to0}
\frac{||\delta{\mathbf x}(t)||}
{||\delta{\mathbf x}(0)||}, 
\end{equation}
that in general depends on the phase space initial position 
${\mathbf x}(0)$ and on the direction in the tangent space 
$\delta{\mathbf x}(0)$. 
If the system is chaotic, the exponential sensitivity to
initial conditions defines a spectrum of $d$ Lyapunov
coefficients $\{\lambda^{(k)}\}_{k=1,2,...d}$ 
coupled in pairs ($d$ being the phase space
dimension), where each element of the pair
is the opposite of the other (symplectic structure).
By the Pesin identity we have that 
$\sum_{\lambda^{(k)}>0}\lambda^{(k)}={\cal K}$ and the
result in \cite{latora_02} states then that
\begin{equation}
\sum_{\langle\lambda^{(k)}\rangle>0}\langle\lambda^{(k)}\rangle
=K,
\label{eq_pesin}
\end{equation} 
where once again $\langle\rangle$ stands for average over
different initial data.

Now, when the largest Lyapunov coefficient vanishes, 
Eq. (\ref{eq_pesin}) 
provides only poor information ($0=0$ , to be precise), which is
not useful for distinguishing, for example, between weak 
chaoticity and regularity. 
In order to address this issue, we generalize the
previous approach in the sense of the nonextensive
generalization \cite{tsallis_02} of the classical (BG)
formalism. 
Specifically, we define the $q$-generalized {\it entropy production} (corresponding to the $q$-generalization of the Kolmogorov-Sinai
entropy rate) as 
\begin{equation}
K_{q_e}\equiv\lim_{t\to\infty}\lim_{W\to\infty}\lim_{N\to\infty}
\frac{\langle S_{q_e}\rangle(t)}{t},
\end{equation}
where the nonextensive entropy $S_{q_e}$ is defined by
\begin{equation}
S_{q_e}\equiv\frac{1-\sum_{i=1}^Wp_i^{\;q_e}}{q_e-1}\;\;\;
(q_e\in{\mathbb R};\;S_1=S_{BG}) \;, 
\label{q_entropy}
\end{equation}
and $e$ stands for {\it entropy}. 
In case of equiprobability, i.e., $p_i=1/W\;(\forall i)$, the
nonextensive entropy takes the form $S_{q_e}=\ln_{q_e}W$, where 
$\ln_qx\equiv(x^{1-q}-1)/(1-q)$  (with $\ln_1 x = \ln x)$ is known in the literature
as the $q$-logarithm \cite{quimicanova}. Notice that its
inverse, the $q$-exponential, is given by 
\begin{equation}
\exp_qx\equiv[1+(1-q)x]^{(1/(1-q))} \;\;(\exp_1x=e^x).
\label{q_exponential}
\end{equation} 
In this paper we will show that, in  
situations where the largest Lyapunov coefficient tends to
zero, 
a regime emerges (that we will discuss in more details later
on) where the sensitivity to initial conditions
takes the form of a $q$-exponential, namely  
\begin{equation}
\xi(t)=\left[1+(1-q_s)\lambda_{q_s} t\right]^{1/(1-q_s)} \equiv \exp_{q_s}(\lambda_{q_s} t)\;,
\end{equation}
with a specific value of
the index $q_s<1$ and of the generalized Lyapunov
coefficient $\lambda_{q_s}$ ($s$ stands for {\it sensitivity}).  
Correspondingly, a single value of $q_e<1$ exists for
which the generalized entropy (\ref{q_entropy}) 
displays a regime of linear increase with time. 
Under these conditions, we study the relation between 
$q_s$ and $q_e$. We will see that, differently
to what happens in the case of strong chaos (where $q_e=q_s=1$), 
the two indices $q_e$ and $q_s$ do not
coincide in general. We discuss the origin of this fact.

\begin{figure}
\begin{center}
\includegraphics[width=\columnwidth,angle=0]{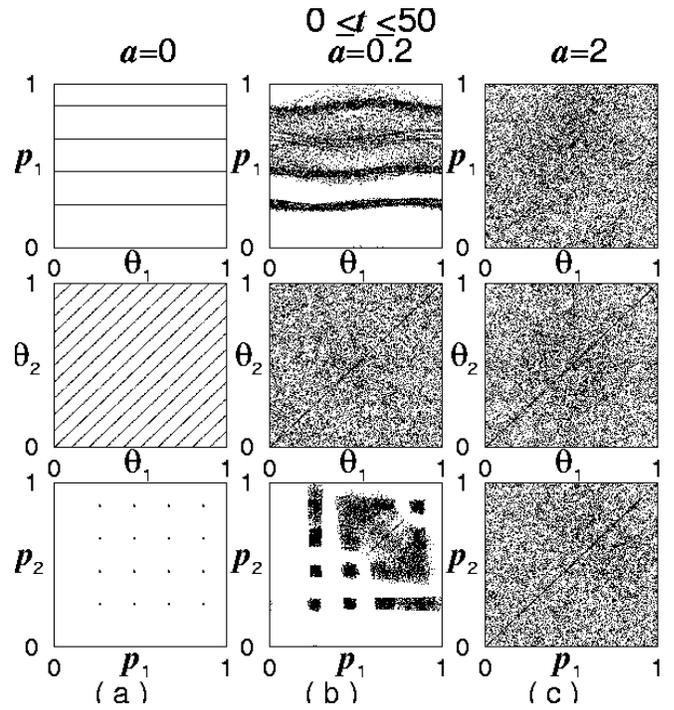}
\end{center}
\caption{\small 
Phase portrait of two symplectically coupled standard
maps, Eq. (\ref{standard_standard}), with $b=0.5$ and
$a_1=a_2\equiv a$. 
Points represent the projection of trajectories on
different planes. We trace an ensemble of $N=4^4$ points,
with a uniform initial distribution, for
$0\leq t\leq50$. 
(a) $a=0$: Integrability;
(b) $a=0.2$: Weak chaos;
(c) $a=2$: Strong chaos.
}
\label{fig_phase_portrait}
\end{figure} 

\section{Sensitivity to initial conditions and entropy
  production} 
\begin{figure}
\begin{center}
\includegraphics[width=\columnwidth,angle=0]{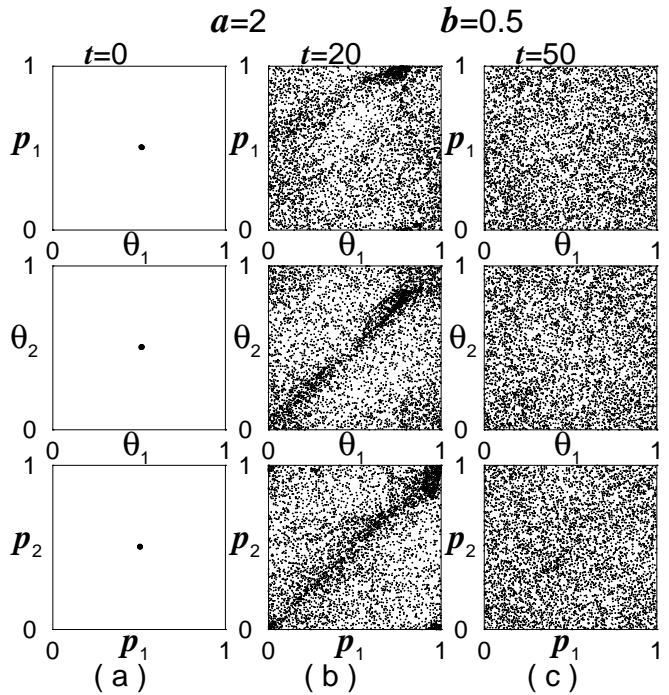}
\end{center}
\caption{\small 
Fixed time  evolution of an initially 
out-of-equilibrium ensemble of
$N=5\times10^3$ copies of the map (\ref{standard_standard}) 
for the strongly chaotic case $a=2$, $b=0.5$.
(a) $t=0$;
(b) $t=20$;
(c) $t=50$.
}
\label{fig_ensemble_strong}
\end{figure}

\begin{figure}
\begin{center}
\includegraphics[width=\columnwidth,angle=0]{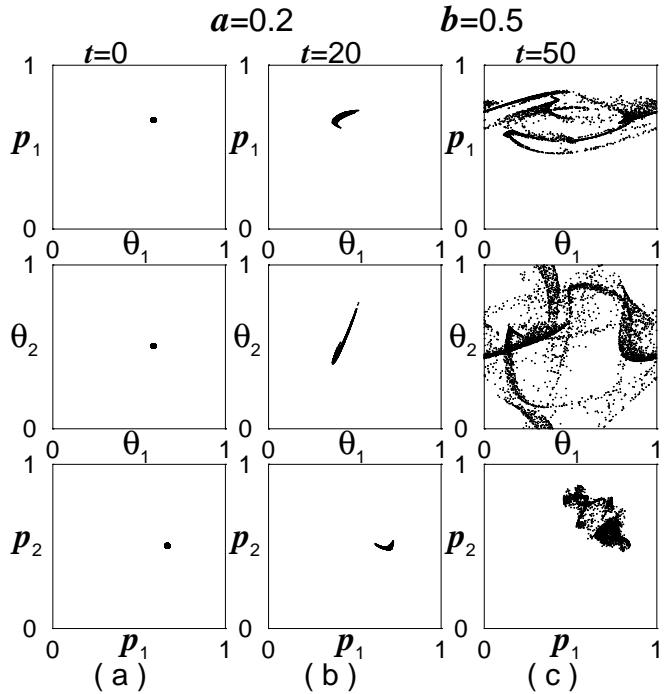}
\end{center}
\caption{\small 
Fixed time evolution of an initially 
out-of-equilibrium ensemble of
$N=5\times10^3$ copies of the map (\ref{standard_standard}) 
for the weakly chaotic case $a=0.2$, $b=0.5$.
(a) $t=0$;
(b) $t=20$;
(c) $t=50$.
}
\label{fig_ensemble_weak}
\end{figure}

We consider a dynamical system with an evolution law given
by two {\it symplectically} coupled standard maps: 
\begin{eqnarray}
\begin{array}{rclr}
\theta_1(t+1)& =& p_1(t+1) + \theta_1(t)+b\;p_2(t+1) & 
({\rm mod}\;1), \\ 
p_1(t+1)&=&p_1(t)+(a_1/2\pi)\sin[2\pi \theta_1(t)]& 
({\rm mod}\;1),\\
\theta_2(t+1)&=&p_2(t+1)+\theta_2(t)+b\;p_1(t+1)& 
({\rm mod}\;1),\\
p_2(t+1)&=&p_2(t)+(a_2/2\pi)\sin[2\pi \theta_2(t)]& 
({\rm mod}\;1),
\end{array}
\label{standard_standard}
\end{eqnarray}
where $\theta_i$, $p_i\in\mathbb{R}$ can be regarded
respectively as an angle and an angular momentum
coordinate, $a_{1},a_{2},b\in\mathbb R$, and 
$t=0,1,2,...\;$. 
Since the system is symplectic $\forall(a,b)$, it immediately
follows that it also is 
{\it conservative}, i.e., the Jacobian determinant of
the map is one ($det\{\partial[\theta_1(t+1),p_1(t+1),\theta_2(t+1),p_2(t+1)]/\partial [ \theta_1(t),p_1(t),\theta_2(t),p_2(t) ]\}=1$). 
For the
sake of definiteness, most part of this paper will deal
with the setting $a_{1}=a_{2}\equiv a$, so that the analysis
will be restricted to the case where the system is symmetric
with respect to exchange $1\leftrightarrow 2$.
If the coupling parameter $b$ vanishes, the two standard maps
decouple.    
For a
generic value of $b$, the system is integrable for $a=0$,
while chaoticity rapidly increases with $|a|$.  
For intermediate values of $|a|$, trajectories in phase
space define complex structures. The phase
space is non-uniform and consists of domains of chaotic
dynamics (stochastic seas, stochastic layers, stochastic
webs, etc.)  and islands of regular quasi-periodic dynamics
\cite{zaslavsky_01}.
To illustrate these behaviors, in
Fig. \ref{fig_phase_portrait}(a-c) we show
the {\it projection} on various planes of the dynamical
evolution of an ensemble of $N=4^4$
points uniformly distributed in phase space at $t=0$,
for $a=2$ (strong chaos), $a=0.2$ (weak chaos), and $a=0$
(integrability).
Reminding the fact that the shadow of a (fractal-like) 
sponge on a wall has a regular geometry, 
the presence of structures as those that appear for $a=0.2$
in the projection of the ensemble over coordinates planes is
a strong hint that even more intricate structures are present in
the phase space itself. 
In order to further characterize the strongly and the weakly
chaotic situations, Figs. \ref{fig_ensemble_strong} and 
\ref{fig_ensemble_weak} 
display, for different {\it fixed} times, the projection of the  
evolution of an initially out-of-equilibrium ensemble, 
respectively for $a=2$ and
$a=0.2$. In all previous figures we have fixed $b=0.5$.

\begin{figure}
\begin{center}
\includegraphics[width=\columnwidth,angle=0]{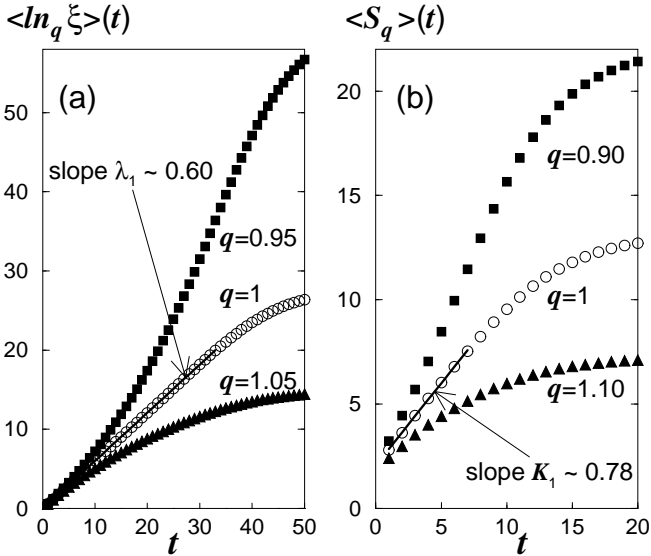}
\end{center}
\caption{\small 
Average value of the logarithm of the sensitivity to initial
conditions (a) and average entropy production (b) for the
map Eq. (\ref{standard_standard}) in a strong chaos regime
($a_1=3$, $a_2=1$ and $b=0.5$). 
The sensitivity to initial conditions is calculated considering
the time evolution of the separation of $10^5$ couple of
initial data randomly distributed in the whole phase space
with an initial separation of order 
$||\delta{\mathbf x}(0)||\simeq 10^{-12}$.
The entropy is calculated dividing the phase space in
$W=30^4$ hypercells and analyzing the spreading of an
ensemble of $N=6\times10^6$ data initially distributed inside
a single hypercell. The calculation was repeated $2000$
times, choosing the position of the initial cell at random
in the whole phase space.
}
\label{fig_latora_baranger}
\end{figure}

Chaotic regions produce exponential separation  of
initially close trajectories.
Before entering in the details of our analysis, let us
consider how the usual, strongly chaotic case is described
inside the nonextensive  formalism, both for the sensitivity
to initial conditions and the entropy production.
To extract the mean value of the largest Lyapunov
coefficient and the mean entropy production rate, 
we analyze respectively the average values
$\langle\ln_q\xi\rangle(t)$ and $\langle S_q\rangle(t)$, 
over different initial conditions. 
Fig. \ref{fig_latora_baranger}, when $q=1$, 
reproduces the results in
\cite{latora_02} for $a1=3$, $a_2=1$
and $b=0.5$ (strong chaos). Only for $q=1$ both the
average logarithm of the sensitivity to initial conditions
and the average entropy display a regime with a {\it linear}
increase with time, before a saturation effect due to the finiteness of
the total number $W$ of (hyper)cells. 
If $q<1$ ($q>1$), the $W \to \infty$ curve is convex (concave) for both
quantities. 
Notice that the slope of the linear
increase of $\langle\ln \xi\rangle(t)$ is smaller than
the slope of $\langle S_{BG}\rangle(t)$, as, since the phase
space is $4$-dimensional, there are {\it two} positive Lyapunov
exponents to be considered in Eq. (\ref{eq_pesin}).

\begin{figure}
\begin{center}
\includegraphics[width=\columnwidth,angle=0]{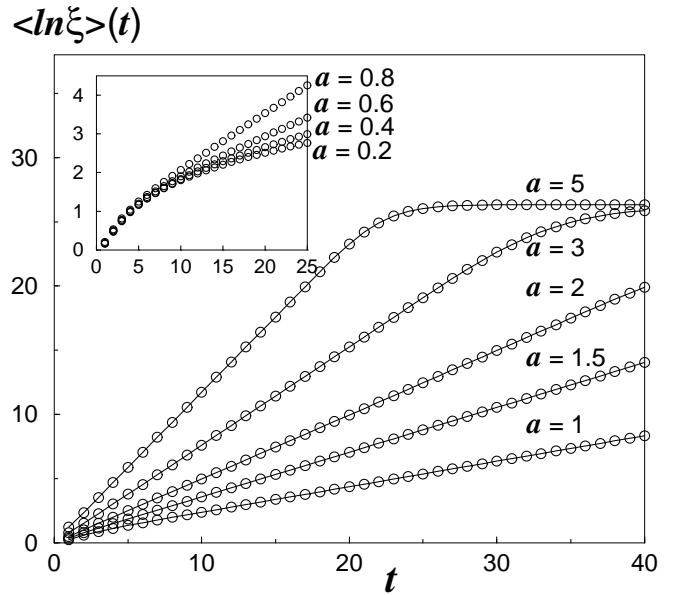}
\end{center}
\caption{\small 
Average value of the logarithm of the sensitivity to initial
conditions for the
map Eq. (\ref{standard_standard}) with $b=0.5$ and various
values of $a$. 
The calculation is the result of the analysis of $10^5$ couple of
initial data randomly distributed in the whole phase space
with an initial separation of order 
$||\delta{\mathbf x}(0)||\simeq 10^{-12}$.
The inset reproduces the same axes of the large figure,
for small values of the ordinate. The solid lines are guides to the eye. 
}
\label{fig_sensitivity_strong}
\end{figure}

We turn now our attention to the weak chaos. 
Fig. \ref{fig_sensitivity_strong} shows that, fixing $b=0.5$
and decreasing now $a_1=a_2\equiv a$, the mean value of the
largest Lyapunov coefficient tends to zero as a consequence
of the fact that dynamics is now more constrained by
structures in phase space. Particularly, for 
$a\lesssim0.8$ an initial phase emerges for which 
$\langle\ln\xi\rangle(t)$  is not linear
(see the inset of Fig. \ref{fig_sensitivity_strong}). 
We refer to this phase as the 
{\it weakly chaotic regime}. 
As $a$ tends to
zero, the crossover time $\tau$ (see its definition in Fig. 6) between this initial regime and the one
characterized by a linear increase of
$\langle\ln\xi\rangle(t)$ increases, so that the weakly chaotic
phase becomes more and more important. 
The sensitivity to initial conditions in correspondence to
this initial regime is power-law. 
In fact, Fig. \ref{fig_sensitivity_weak} exhibits that, for
specific values $q_s(a)<1$, $\langle\ln_{q_s}\xi\rangle(t)$
grows {\it linearly} during this regime and starts later to be
convex when the sensitivity becomes exponential, after the
crossover. For $a$ approaching zero, the crossover time
between this initial regime and the strongly chaotic regime
(characterized by a linear increase of
$\langle\ln\xi\rangle(t)$) {\it diverges} (see inset of
Fig. \ref{fig_sensitivity_weak}). 
Note that $q_s$ {\it depends} on $a$. 
For integrability we have, as expected,
$q_s(0)=0$.

\begin{figure}
\begin{center}
\includegraphics[width=\columnwidth,angle=0]{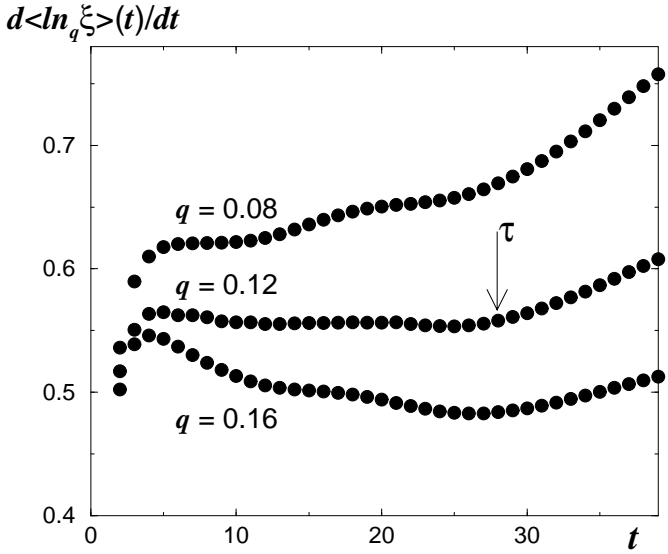}
\end{center}
\caption{\small 
Time dependence of the time derivative of the average $q$-logarithmic sensitivity for $a=0.05$ and $b=0.5$ for typical values of $q$. We verify that $q_s \simeq 0.12$ and $\lambda_{q_s} \simeq 0.56$. The arrow indicates the value of $\tau$ .
}
\label{fig6}
\end{figure} 

\begin{figure}
\begin{center}
\includegraphics[width=0.9\columnwidth,angle=0]{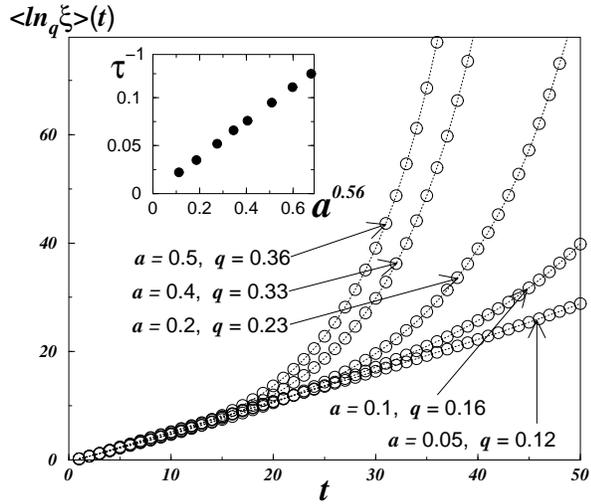}
\end{center}
\caption{\small 
Average value of the $q_s$-logarithm of the sensitivity to initial
conditions for the
map Eq. (\ref{standard_standard}) with $b=0.5$ and various
values of $a$. 
The calculation is the result of the analysis of $10^8$ couple of
initial data randomly distributed in the whole phase space
with an initial separation of order 
$||\delta{\mathbf x}(0)||\simeq 10^{-12}$.
$q_s$ changes with $a$ (see also
Fig. \ref{fig_qs_qe_weak}) and it was determined by a
constant derivative condition during the weakly chaotic
regime. The dotted lines are guides to the eye. 
In the inset, inverse of the crossover time vs $a^{0.56}$.   
}
\label{fig_sensitivity_weak}
\end{figure} 

\begin{figure}
\begin{center}
\includegraphics[width=\columnwidth,angle=0]{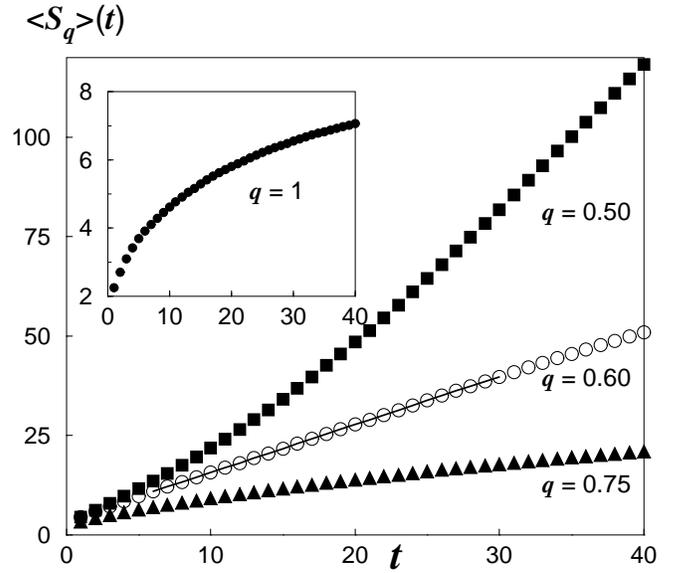}
\end{center}
\caption{\small 
Average entropy production for the
map Eq. (\ref{standard_standard}) in a weak chaos regime
$b=0.5$ and $a=0.2$. 
Phase space was divided into $W=30^4$ hypercells and the
spreading of an ensemble of $N=6\times10^6$ initial data was
analyzed. The calculation was repeated $3000$
times, choosing the position of the initial cell at random
in the whole phase space. We verify that $q_e\simeq0.6$ and
$K_{q_e}\simeq1.2$; 
$q_e$ changes with $a$ (see also
Fig. \ref{fig_qs_qe_weak}) and it was determined by a
constant derivative condition during the weakly chaotic
regime. The straight line is the result of a linear interpolation. 
}
\label{fig_entropy_weak}
\end{figure}

The weakly chaotic regime can also be detected by an
analysis of the entropy production, performed as described in
section \ref{section_introduction}. 
In Fig. \ref{fig_entropy_weak} 
we take as an example the case $b=0.5$ and $a=0.2$. 
For the weakly chaotic regime the result is similar to 
Fig. \ref{fig_latora_baranger}(b), but the value for which
the entropy grows linearly is now $q_e \simeq 0.6$ instead of
$q_e=1$. 

\begin{figure}
\begin{center}
\includegraphics[width=\columnwidth,angle=0]{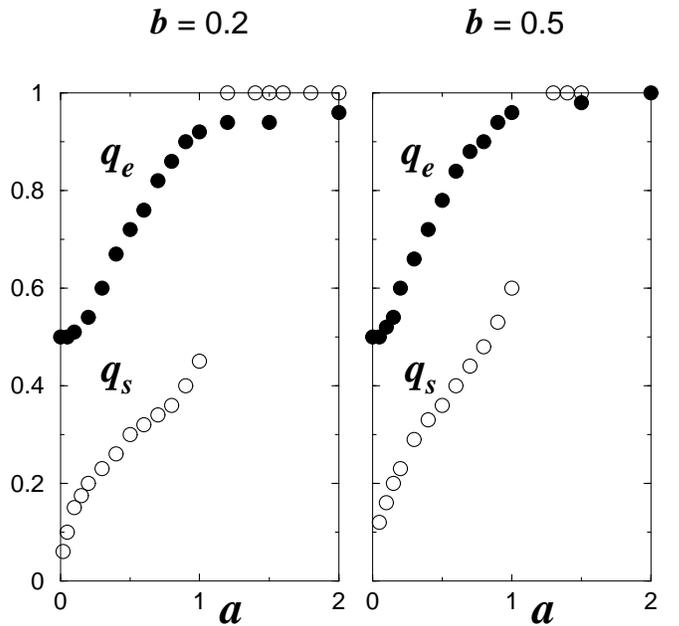}
\end{center}
\caption{\small 
Behavior of $q_s$ and $q_e$ as a function of $a$, for 
$b=0.2$ e $b=0.5$. The two parameters coincide only in the
case of strong chaos ($|a|>>1$), when $q_s=q_e=1$.
}
\label{fig_qs_qe_weak}
\end{figure}

Fig. \ref{fig_qs_qe_weak} synthesizes the behavior of $q_s$
and $q_e$ for two different fixed coupling constants, 
namely $b=0.5$ and $b=0.2$.
At variance with what happens in the strongly chaotic regime,
in the weakly chaotic one $q_s$ and $q_e$ are {\it not} the same. 
In the next section we discuss the origin of this difference.

Another numerical result is that,  while the the generalized
Lyapunov coefficient appears to be almost independent from
$a$ ($\lambda_{q_s}=0.57$ and $\lambda_{q_s}=0.50$ for  
$b=0.20$ and $b=0.50$ respectively), the generalized
Kolmogorov-Sinai entropy rate $K_{q_e}$ exhibits, within our
numerical precision, a slight
decrease 
as $a$ increases. We do not have
yet a clear explanation for this observation.

\section{Discussion}
To understand the origin of the discrepancy, in the
weakly chaotic regime, between $q_s$ and $q_e$ some simple
geometric considerations are in order. 

Firstly, we notice the following property of the $q$-logarithmic
function:
\begin{equation}
\ln_qW^\alpha=\alpha\ln_{q^\prime}W \;,
\label{q_log_prop_1}
\end{equation}
where $\alpha\in\mathbb R$ and $q$, $q^\prime$ are related by 
\begin{equation}
1-q=\frac{1-q'}{\alpha}.
\label{q_log_prop_2}
\end{equation}
Notice that, for all $\alpha>1$, $q^\prime=1$ implies $q=1$, whereas $q^\prime<1$ implies $q^\prime < q <1$ .

Secondly, for simplicity, we consider the case of a
$2$-dimensional phase space where the dynamical evolution is
{\it symmetric} with respect to the exchange of the coordinates. 
Let us call $W^{(2)}=W^{(1)}\otimes W^{(1)}$ the total 
number of bidimensional cells composed by a partition of $W^{(1)}$
intervals in each coordinate. 
We suppose to set a far-from-equilibrium initial ensemble
inside a single $2$-dimensional 
cell and that the dynamical evolution of
each trajectory in the ensemble implies a spreading of the
ensemble in phase space. 
Let $W^{(1)}(t)$ and $W^{(2)}(t)$ be respectively the number
of occupied \mbox{$1$-dimensional} 
intervals in one of the two
coordinate axes and the number of occupied
\mbox{$2$-dimensional} cells, at time $t$ ($W^{(1)}(0) = W^{(2)}(0)=1$). 
The relation between $W^{(1)}(t)$ and $W^{(2)}(t)$  depends
on the details of the dynamical evolution. 
Two limiting cases are 
\begin{equation}
W^{(2)}(t)\propto W^{(1)}(t)\;\;\;{\rm and}\;\;\;
W^{(2)}(t)\propto [W^{(1)}(t)]^2.
\end{equation}
The former is realized for instance when there is a
predominant direction 
along which the ensemble stretches, so that the dynamical
evolution of the ensemble produces filaments in the
$2$-dimensional phase space and it is essentially unidimensional. 
The latter happens for example when the dynamical evolution
in the two coordinates is decoupled, so that $W^{(2)}(t)$ is
simply the Cartesian product $W^{(1)}(t)\otimes W^{(1)}(t)$ .
Now, if we suppose that $\ln_{q_s}W^{(1)}(t)\propto t$, for a
specific value of the parameter $q_s$ 
(we are using the notation $q_s$ consistently with the fact
that $\xi$ represents the
growth of a unidimensional arc), by means of
Eqs. (\ref{q_log_prop_1}) and (\ref{q_log_prop_2}) we have that
$\ln_{q_e}W^{(2)}(t)\propto t$ for the values
\begin{equation}
q_e=q_s\;\;\;{\rm and}\;\;\;
q_e=1-\frac{1-q_s}{2},
\end{equation}
respectively, for the two previous limiting cases (in Eq. (11)).
Notice that $q_s=1\Rightarrow q_e=1$ for both cases. 
Of course, Eq. (12) can be straightforwardly generalized for arbitrary
dimensions of the phase space. 

\begin{figure}
\begin{center}
\includegraphics[width=\columnwidth,angle=0]{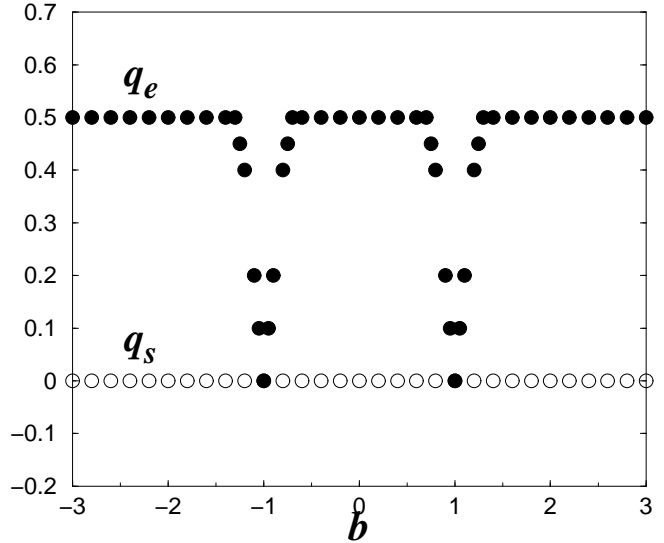}
\end{center}
\caption{\small 
Behavior of $q_s$ and $q_e$ as a function of $b$, for the
integrable case $a=0$. The two indices coincide only
for the special condition $b=\pm1$ (see text for
details), when $q_s=q_e=0$. For almost all other values of $b$ we verify $1-q_e=(1-q_s)/2$ .
}
\label{fig_qs_qe_integrable}
\end{figure} 

\begin{figure}
\begin{center}
\includegraphics[width=\columnwidth,angle=0]{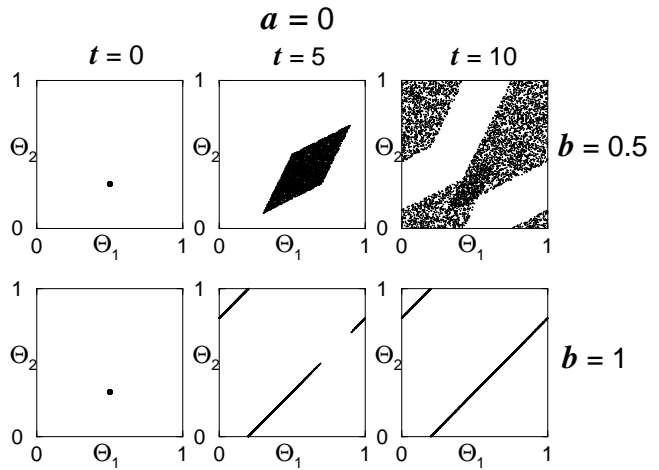}
\end{center}
\caption{\small 
Fixed time evolution of an initially 
out-of-equilibrium ensemble of
$N=5\times10^3$ copies of the map (\ref{standard_standard}) 
for the integrable case $a=0$, $b=0.5$ (first row) and
$b=0.2$ (second row). We display the projection of the
ensemble on the plane $(\theta_1,\theta_2)$.
}
\label{fig_ensemble_integrable}
\end{figure} 

This geometrical analysis applies to the problem of the
entropy production, strictly speaking, only if the ensemble
evolves according to a uniform distribution in phase space
(equiprobability), so that the value of the generalized
entropy is given by $\ln_qW$.  
Nonetheless, we will see now that the same analysis is useful to understand
the dynamical behavior of the model that we are studying. 
We start by considering the effect of the coupling term $b$
in the special case $a=0$, that corresponds to
integrability.  
Fig. \ref{fig_qs_qe_integrable} 
represents the behavior of $q_s(b)$ and $q_e(b)$ for
this case. As expected, $q_s$ is zero, thus displaying
that the sensitivity to initial conditions increases linearly. 
On the other hand, the behavior of $q_e$ is understood if we
analyze in some details the dynamics. We have that $p_1$ and
$p_2$ are just conserved along any trajectory and all the
interesting dynamical evolution take place in the
$(\theta_1,\theta_2)$-plane by means of the iteration laws
\begin{eqnarray}
\theta_1(t+1)&=&p_1(0)+\theta_1(t)+b\;p_2(0),\\
\theta_2(t+1)&=&p_2(0)+\theta_2(t)+b\;p_1(0).\nonumber
\end{eqnarray}
As it is apparent, the two coordinates are decoupled and for
$b\neq\pm1$ the growth of the ensemble is $2$-dimensional as
confirmed by Fig. \ref{fig_ensemble_integrable} (first row). 
Under these conditions, we have 
$W^{(4)}(t)\propto W^{(2)}(t)\propto [W^{(1)}(t)]^2$, so
that, applying the previous analysis, we obtain $q_e=0.5$.
For $b=\pm1$ the maps further degenerates, since now we have
one more conserved quantity, i.e.,  
\begin{equation}
\theta_1(t+1)\mp\theta_2(t+1)=\theta_1(t)\mp\theta_2(t).
\end{equation}
This implies 
$W^{(4)}(t)\propto W^{(2)}(t)\propto W^{(1)}(t)$
(see Fig. \ref{fig_ensemble_integrable}, second row), and we
obtain $q_e=q_s=0$.
Around $b=\pm1$ there is then a transition from $q_e=0.5$
to $q_e=0$, as it is exhibited in 
Fig. \ref{fig_qs_qe_integrable}.  

\begin{figure}
\vspace{1cm}
\begin{center}
\includegraphics[width=0.98\columnwidth,angle=0]{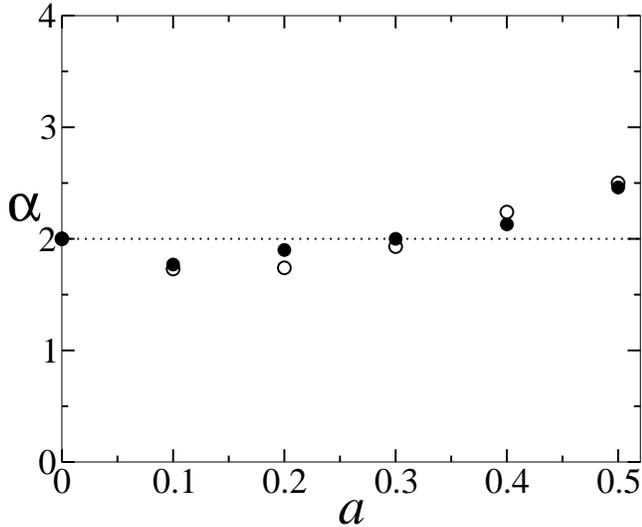}
\end{center}
\caption{\small 
Behavior of $\alpha$ as a function of $a$, for 
$b=0.2$ (open circles) and $b=0.5$ (full circles). 
For a large part of the weakly chaotic region $\alpha$ is
almost constant and equal to $2$.
The line is a guide to the eye.
We do not show the data for $a>0.5$ because the time interval
within which we define $q_s$ and $q_e$ is not large enough
to provide a confident value of these quantities.
}
\label{fig_alpha}
\end{figure} 

We can consider now the generic weakly chaotic case ($0<|a|<<1$). 
In Fig. \ref{fig_alpha} we estimate the (below defined) factor $\alpha$ that connects
the $4$-dimensional analysis performed by the entropy to the
one-dimensional inspection provided by the sensitivity to
initial conditions, 
by means of the relation (see Eq. (\ref{q_log_prop_2})) 
\begin{equation}
\alpha \equiv \frac{1-q_s}{1-q_e}.
\label{relation_alpha}
\end{equation}
For both $b=0.5$ and $b=0.2$, small positive values of $a$ yield $\alpha \simeq 2$ , as it is for the integrable case $a=0$. This exhibits an essentially
$2$-dimensional growth of the portion of the phase
space occupied  by the ensemble. 

At this point we can advance a conjecture for the more
general case of a phase space that is not symmetric under
the interchange of coordinates. This conjecture  will allow for a better
understanding of the reason of the discrepancy between 
$q_s$ and $q_e$. 
In the case of strong chaos in a symplectic nonlinear system, the growth of the hypervolume
containing the initially out-of-equilibrium ensemble is
essentially $d/2$-dimensional ($d$ being the phase space
dimension of the system). In fact, if  
$\{\lambda^{(k)}\}_{k=1,2,\ldots,d/2}$  with 
$\lambda^{(k)}>0\;\forall k$ is the spectrum of 
{\it positive} Lyapunov coefficients, and we call 
$\xi^{(k)}$ the sensitivity to initial conditions associated
to the direction defining $\lambda^{(k)}$, we have
\begin{eqnarray}
W^{(d)}(t)&\propto&W^{(d/2)}(t)\\
&\propto&\xi^{(1)}(t)\;\xi^{(2)}(t)\;\cdots\;\xi^{(d/2)}(t)\\
&\propto&
\exp(\lambda^{(1)}t)\;\exp(\lambda^{(2)}t)
\;\cdots\;\exp(\lambda^{(d/2)}t)\\
&\propto&\exp\left(\sum_{k=1}^{d/2}\lambda^{(k)}t\right).
\end{eqnarray}
We remind that $W^{(d)}(t)$ and $W^{(d/2)}(t)$ in Eq. (16) respectively are the numbers of occupied $d$-dimensional and $d/2$-dimensional hypercells. 
Taking the logarithm of relation (19) we obtain essentially the
Pesin equality, namely 
\begin{equation}
K=\sum_{k=1}^{d/2}\lambda^{(k)},
\label{pesin_repeated}
\end{equation}
where $\ln W^{(d)}(t) \simeq Kt$.

If we now assume that the sensitivities to initial conditions
along the various directions of the expansion are 
$q^{(k)}$-exponential power-laws (which implies that the largest Lyapunov coefficient
vanishes) 
\begin{equation}
\xi^{(k)}\propto t^{1/(1-q^{(k)})},
\end{equation}
the previous relations (16-19) transform, for very long times, into  
\begin{eqnarray}
W^{(d)}(t)&\propto&W^{(d/2)}(t)\\
&\propto&\xi^{(1)}(t)\;\xi^{(2)}(t)\;\cdots\;\xi^{(d/2)}(t)\\
&\propto&
t^{1/(1-q^{(1)})}\;t^{1/(1-q^{(2)})}
\;\cdots\;t^{1/(1-q^{(d/2)})}\\
&\propto&t^{\sum_{k=1}^{d/2}1/(1-q^{(k)})},
\end{eqnarray}
and, since $\ln_{q_e} W^{(d)}(t)$ is linear with time for
$q_e$ (i.e., $S_{q_e} \propto    [W^{(d)}(t)]^{1-q_e} \propto t$, hence $W^{(d)}(t) \propto t^{1/(1-q_e)}$) we obtain
\begin{equation}
\frac{1}{1-q_e}=\sum_{k=1}^{d/2}\frac{1}{1-q^{(k)}}.
\label{standard_2_pesin_gen}
\end{equation}
Note that, for $q^{(k)}<1\;\forall k$, we have
$q_e>q^{(k)};\forall k$. 
Note also that for 
$q^{(k)}=q_s\;\forall k$ Eq. (\ref{standard_2_pesin_gen})
reduces to Eq. (\ref{relation_alpha}) identifying
$\alpha\equiv d/2$.

Since Eq. (\ref{standard_2_pesin_gen}) relates an information associated with the entropy production to an
information associated with the sensitivity to initial
conditions, it plays, for weakly chaotic dynamical systems, the role played by the Pesin equality for strongly chaotic ones. 
Working out a generalization of traditional methods for the
calculation of the complete spectrum of the Lyapunov
coefficients (e.g., \cite{benettin_01}), 
it might be possible to numerically verify the validity of Eq. (\ref{standard_2_pesin_gen})
\cite{baldovin_0x}.

Eqs. (\ref{pesin_repeated}) and (\ref{standard_2_pesin_gen})
can be unified as follows ($t>>1$)
\begin{equation}
\exp_{q_e}(K_{q_e}t)\simeq\prod_{k=1}^{d/2}\left[
\exp_{q^{(k)}}(\lambda_{q^{(k)}}^{(k)} \;t)\right].
\label{q_pesin_conjectured}
\end{equation}
This relation hopefully is, for large classes of dynamical systems yet to be qualified, a correct conjecture. If so it is, then it certainly constitutes a powerful relation. Indeed, let us summarize some of its consequences: 

(i) If the system is strongly chaotic, i.e., if it has positive Lyapunov exponents, we have $q_e=q_s=1$ and Eq. (20) holds;

(ii) If the system is weakly chaotic (hence its largest Lyapunov exponent vanishes), we have that Eq. (26) holds; 

(iii) If the system is weakly chaotic and there is only {\it one} dimension within which there is mixing, then,  not only $q_e=q_s$, but also
\begin{equation}
K_{q_e}=\lambda_{q_s} \;,
\end{equation}
as already known for unimodal maps such as the logistic one \cite{baldovin_01,baldovin_02};

(iv) If the system is weakly chaotic and there is isotropy in the sense that $q^{(k)}= q_s$ ($\forall k=1,...,d/2$), we have that
\begin{equation}
\frac{1}{1-q_e}=\frac{d/2}{1-q_s} \;.
\end{equation}
The special case $q_s=0$ yields
\begin{equation}
q_e=1-\frac{2}{d} \;.
\end{equation}
It is suggestive to notice that, discussing a Boltzmann transport equation concerning a fluid model with Galilean-invariant Navier-Stokes equations in a $d_{NS}$-dimensional Bravais lattice, Boghosian et al \cite{boghosianetal} obtained $q_e=1-2/d_{NS}$, and that, for a lattice Lotka-Volterra $d_{LV}$-growth model, Tsekouras et al \cite{tsekourasetal} obtained $q_e=1-1/d_{LV}$ ($NS$ and $LV$ stand respectively for {\it Navier-Stokes} and {\it Lotka-Volterra}).

\section{Conclusions}
We have discussed the (average) sensitivity to initial
conditions and entropy production for a $4$-dimensional 
dynamical system
composed by two symplectically and symmetrically coupled
standard maps, focusing on phase space configurations
characterized by the presence of complex (fractal-like)
structures. 
Under these conditions, coherently with previous
$2$-dimensional observations \cite{baldovin_03}, we have
detected the emergence of long-lasting regimes
characterized by a power-law sensitivity to initial
conditions, whose duration diverges when the map parameters
tend to the values corresponding to integrability.
While the classical BG formalism characterizes these
anomalous regimes only trivially by means of the Pesin
equality \cite{pesin} (we have $K=0$ and $\lambda^{(k)}=0\;\forall k$
in Eq. (\ref{eq_pesin})),
the nonextensive formalism \cite{tsallis_02}, 
through a generalization of the
Pesin equality (see also \cite{tsallis_01,baldovin_02}),
provides a meaningful nontrivial description
for these regimes. 
Specifically, we have shown that, during these anomalous
regimes (here called weakly chaotic) 
the average sensitivity to initial conditions is a
$q_s$-exponential power-law (\ref{q_exponential}), 
with $q_s<1$, and that the
corresponding entropy production is (asymptotically) {\it finite} only for a generalized $q_e$-entropy
(\ref{q_entropy}), with $q_e<1$. 
We have discussed the relation between $q_s$ and $q_e$, both
numerically and analytically and we propose an appealing 
generalization of the Pesin equality, namely 
Eq. (\ref{q_pesin_conjectured}).
By means of the difference between $q_s$ and $q_e$ we
obtain useful information about the dimensionality which
is associated to the spreading of the initially
out-of-equilibrium dynamical ensemble. 
If this dimensionality is equal to one, we have the
result $q_e=q_s$ and $K_{q_e}=\lambda_{q_s}$, as already 
conjectured \cite{tsallis_01} and proved
\cite{baldovin_02} for the edge of chaos of unimodal maps. 

The present results concern low-dimensional conservative
(Hamiltonian) maps, but the scenario we have described here
may serve as well for the discussion of analogous effects
arising when many maps are symplectically coupled (see,
e.g., \cite{baldovin_05}), and even for the case of many-body
interacting Hamiltonian systems. Short-range interactions would typically yield strong chaos, and long-range interactions may typically lead to weak chaos (consistently with the results in \cite{anteneodo,cabral}).
In fact, for isolated many-body long-range-interacting classical Hamiltonians there are vast classes of initial conditions  for which metastable (or quasi-stationary) states are currently observed, later on followed by a crossover to the usual Boltzmann-Gibbs thermal equilibrium (see, for instance, \cite{rapisardaetal}). The duration of the metastable state {\it diverges} with the number $N$ of particles in the system, in such a way that the limits $N \to \infty$ and $t\to\infty$ do {\it not} commute. Since it is known that such systems may have vanishing Lyapunov spectrum, it is allowed to suspect that the scenario in the metastable state is similar to the weakly chaotic one described in the present paper, whereas the crossover to the BG equilibrium corresponds, in the system focused on here, to the crossover to the $q_e=q_s=1$, strongly chaotic regime. These and other crucial aspects can in principle be verified, for instance, on systems with very many (and not only two, as here) symplectically coupled standard maps. They would provide insighful information about macroscopic systems and their possible universality classes of nonextensivity (characterized by index(es) $q$). Efforts along this line are surely welcome.  

\section*{Acknowledgments}

We acknowledge useful discussions with C. Anteneodo, E.P. Borges, L.G. Moyano, A. Rapisarda and A. Robledo, as well as partial financial support by Pronex/MCT, Faperj, Capes and CNPq (Brazilian agencies).

\end{document}